\documentstyle{camera}
\newcommand{\AmS}{{\protect\the\textfont2
  A\kern-.1667em\lower.5ex\hbox{M}\kern-.125emS}}

\newcommand{\beq}{\begin{equation}}
\newcommand{\eeq}{\end{equation}}
\newcommand{\bea}{\begin{eqnarray}}
\newcommand{\eea}{\end{eqnarray}}

\def\dm2{\Delta m^2}
\def\sq2{sin^2(2\Theta)}

\input epsf

\begin{document}

%%%%%%%%%%%%%%%%%%%%%%%%%%%%%%%%%%%%%%%%%%%%%%%%%%%%%%%%
% The title, all uppercase; if you want to split it in
% two or more lines, put a \\ macro at each line break
% example:
%   \title{TITLE: FIRST LINE\\ SECOND LINE}
%
\title{ Cosmic Rays Astrophysics and  Neutrino  Astronomy beyond and beneath the  Horizons }

 \author{D. Fargion}
%%%%% \author[label1]{, M. Grossi}
%%%%%  \author[label1]{, M. De Santis}
%%%%%  \author[label1]{ P.G. De Sanctis Lucentini}
%%%%%  \author[label1,label2]{, M.Iori}
%%%%%  \author[label1]{, A. Sergi}
%%%%%   \author[label1]{, F.Moscato}
%%%%%% \address[label1]{Dep. of Physics, University of Rome "La Sapienza"}
%%%%%\address[label1]{Physics Department , INFN, Universit\'a di Roma  "La Sapienza", Pl. A. Moro
%%%%%2, 00185, Rome,  Italy}
\organization{ Physics Department, University of Rome "La
Sapienza" \\
and INFN Rome1, Italy}

%\author[label1]{M. Grossi}

%\address{}

\maketitle

\begin{abstract}
 Modern Terrestrial Cerenkov Telescopes  and Array Scintillators facing the Horizons may  soon reveal
far  Cosmic Rays  or  nearer PeVs-EeVs Neutrino Showers Astronomy.
Also UHE neutrino interactions in air, leading to Horizontal
Showers, may take place through several channels: the main Glashow
resonant $\overline{\nu}_{e}-e \rightarrow W^-$ one, the charged
nuclear interactions ${\nu}_e + N \rightarrow e^- + X $,
${\nu}_{\mu} + N \rightarrow \mu + X $, ${\nu}_{\tau} + N
\rightarrow \tau + X $,
 and the neutral current events  ${\nu}_e + N \rightarrow {\nu}_e + X $, ${\nu}_{\mu} + N \rightarrow
{\nu}_{\mu} + X $, ${\nu}_{\tau} + N \rightarrow {\nu}_{\tau}+ X
$. Analogous events occur also for $\bar{\nu}$-nucleon events.
 These interactions are producing hadronic or electromagnetic
showers at the far horizons and their correlated secondaries
Cerenkov flashes; additional double showering in air may occur by
rare tau decay in flight. Comparable interactions ${\nu}_{\tau} +
N \rightarrow \tau + X $ are producing, inside the Earth, the
Horizontal, but Up-going ultra-relativistic tau, whose decay in
flight is source of $\tau$ Air Showers (Uptaus and HorTaus);
Horizontal Tau Air-Showers are rarely (one-two order of magnitude)
enhanced by highest mountain chains at horizons, as in AUGER
experimental site; their hadronic or electromagnetic traces are
preferentially at PeV-EeV energies; their tracks are amplified by
their muon-gamma bundles whose signals are above the random noises
and better disentangled by Cerenkov flashes in time coincidence.
More exotic UHE SUSY interactions $\chi^o + e\rightarrow \tilde{e}
\rightarrow\chi^o + e $ at tens PeVs-EeV energy may blaze at
Horizons  if $\tilde{e}_R$ has a mass comparable with gauge boson
$W^-$ one ( within a top-down model for UHECR). Anyway common High
Energy Cosmic rays at PeVs-EeVs energy $must$ traces their
presence in Magic-like Cerenkov Telescopes at horizons; every
night these telescopes may observe tens of clustered C.R. showers
at $87^o-90^o$ zenith angle by  the far precursor Cerenkov flashes
and later (micro-second) muon rings (hitting inside the telescope)
and-or decay (into electron) in flight leading to near-by
mini-gamma showers. They may be a first probe to calibrate a Novel
  rare Neutrino Astronomy below the horizons. Present Magic Telescopes
  facing the Horizons  are, at equal given time,  almost comparable with the
  Amanda underground  neutrino detector.
\end{abstract}

% main text

\section{Cosmic Ray Flashes above the  Earth edges}

Ultra High Energy and High Energy Cosmic Rays (UHECR) Showers
(PeVs - EeVs band energy) born at high altitude in the atmosphere
may be detected by telescopes such as Magic provided that they are
pointed \emph{above the horizon} .
 These Cerenkov telescopes, set in a crown circular array toward the Horizons, may be correlated by an analogous
  scintillator  ones able to trace muons and electromagnetic shower
  particles.  The earliest gamma and Cerenkov lights produced
   by any down-ward horizontal, zenith angle $(85^o-90^o$ ),
   cosmic rays (CR), observed by a high  ($> 2$ km) mountain
 (whose C.R. nature is mainly hadronic), while  crossing  the atmosphere
  are severely absorbed by the thick horizontal air column depth ($10^4 - 5 \cdot 10^4$ $ g \cdot
cm^{-2})$. However the most energetic C.R. blazing Cerenkov shower
 must survive and also revive during their propagation. Indeed
one has to expect that these CR showers contain also a diluted but
more penetrating component made of muon bundles, which can be
detected in two ways: a) by their Cerenkov lights emitted after
they decay into electrons near the Telescope; b) by the same muons
hitting inside the Telescope,  blazing  ring or arc by the same
muon Cerenkov lights. The latter muon bundles  are less abundant
compared to the peak gamma bundles  produced in the shower
(roughly $10^{-3}$ times lower). They are mostly produced  at an
horizontal distance of $100-500$ km from the observer (for a Magic
site placed at 2.2km above the sea level and a  zenith
  angle of  $85^{\circ}- 91.5^{\circ}$).
Therefore these hard muon bundles (each one at tens-hundred GeV
energy) might spread in large areas of tens - hundred $km^2$ while
they travel towards the observers. They are partially split and
bent by the local geomagnetic field,  along an axis orthogonal to
the field and along their propagation direction; a fraction of
them may decay into electrons producing luminous mini-optical
Cerenkov flashes. The geo-magnetic spread of the shower leads to
an early aligned Cerenkov blaze whose shape (a twin split shower)
and inclination may probe the magnetic field "polarization" and
the CR origination. Such a characteristic signature may be
detected by the largest gamma telescope arrays as Hess, Veritas,
or the forthcoming stereoscopic version of Magic. We argue that
their Cerenkov flashes, either single or clustered, must take
place, at detection threshold, at least tens or hundreds times a
night for Magic-like Telescope pointing toward horizons at a
zenith angle between $85^{\o}$ and $90^{\o}$. Their easy
"guaranteed" discover may offer, we believe, a new tool in CR and
UHECR detection. At zenith angle between $85^{\o}$ and $80^{\o}$
by an accurate statistics there is the possibility to disentangle
the cosmic ray spectra and composition at PeV-EeV energy windows.
Their primary hadronic signature might be hidden by the distance
but its late tail may arise in a new form by  its secondary
muon-electron-Cerenkov  of electromagnetic showering.

 Moreover    a rarer   but  more exciting PeV - EeV
Neutrino ${\nu_{\tau}}$ Astronomy (whose flux is suppressed by
nearly three-four orders of magnitude respect to CR one) may arise
\emph{below the horizon} with the Earth-Skimming Horizontal Tau
Air-Showers (HorTaus); below the horizons one is not awaiting C.R.
shower, because Earth opacity,  at all. These UHE Taus are
produced inside the Earth Crust by the primary UHE incoming
neutrino ${\nu}_{\tau}$, $\overline{\nu}_{\tau}$, and they are
generated mainly by the muon-tau neutrino oscillations from
galactic or cosmic sources, \cite{Fargion1999}\cite{Fargion
2002a}, \cite{Feng2002}, \cite{Bertou2002}, \cite{Fargion2004}.

Moreover we expect also that just above (one-two degree) or below
the horizon edge (half a degree), within a distance of a few
hundreds of km, fine-tuned $\overline{\nu}_e$-$e\rightarrow
W^-\rightarrow X$ Glashow resonance at $6.3$ PeV might  be
detectable as horizontal air-showers; they are nearly two-three
orders of magnitude below CR ones (for comparable incoming
fluxes). The W main hadronic ($2/3$) or leptonic and
electromagnetic ($1/3$) signatures  may be well observed and their
rate may be used to calibrate a new horizontal neutrino
multi-flavour Astronomy \cite{Fargion 2002a}. Again we argue that
such a signature of nearby nature (respect to most far away ones
at same zenith angle of hadronic nature) would be better revealed
by  a Stereoscopic twin telescope such as Magic or a Telescope
array like Hess, Veritas.

       Additional Horizontal flashes  might arise
    by Cosmic UHE $\chi_o + e \rightarrow  \widetilde{e}\rightarrow \chi_o +
    e$ electromagnetic showers  within most SUSY models, if UHECR are born in topological
   defect decay or in their annihilation, containing a relevant component of SUSY
   particles. The UHE $\chi_o + e \rightarrow  \widetilde{e}\rightarrow \chi_o +
    e$ behaves (for light $\widetilde{e}$ masses around Z boson ones)
    as the Glashow  resonance peak \cite{Datta04}.
The total amount of air inspected within the characteristic field
of view of  MAGIC ($2^{\o} \cdot 2^{\o}$) at the  horizon ($360$
km.) corresponds to a (water equivalent) volume-mass larger than $
V_w\simeq 44 km^3$. However their detectable beamed volume
corresponds  to a  narrower thinner volume   $ V_w\simeq 1.36
\cdot 10^{-2}$ $km^3$,  yet  comparable to the  present AMANDA
confident volume  (for Pevs $\overline{\nu}_e$-$e\rightarrow
W^-\rightarrow X$   and EeVs  ${\nu}_{\tau}$,
$\overline{\nu}_{\tau} + N \rightarrow \tau\rightarrow $ showers).

\section{ Flashes  by Prompt Showers and Muons}

  The ultrahigh energy cosmic rays (UHECR) have been studied
   in the past mainly through their secondary particles ($\gamma$, $e^\pm$, $\mu^\pm$)
  collected vertically in large  array detectors on the ground. UHECRs are rare events, however the multiple cascades occurring at high
  altitudes where the shower usually takes place, expand and amplify  the signal detectable on the ground.
  On the other hand, at the horizon the UHECR  are hardly observable (but also rarely
  searched).  They are diluted both by  the larger  distance they have to cover and
by the atmosphere opacity suppressing exponentially their
electromagnetic  secondaries (electron pairs and gammas); also
their optical Cerenkov emission is partially  suppressed by the
horizontal air opacity.
   However this suppression acts also as a useful filter
  leading to the selection of higher CR events. Their Cerenkov lights
  will be scattered and partially transmitted ($1.8\cdot 10^{-2}$ at $551$ nm., $6.6\cdot 10^{-4}$ at $445$ nm.)
  depending on the exact zenith angle and the seeing: assuming an average suppression factor -  $5\cdot
  10^{-3}$ for the air opacity and $10^{-3}$ for the $30$ times larger distances,
    the  nominal Magic threshold at $30$ GeV does
      correspond to a hadronic shower coming from the horizon
  with an  energy above $E_{CR}\simeq 6 $ PeV.
     Their primary flux may be estimated considering
     the known cosmic ray fluxes at the  same energy on the top of the atmosphere (both protons
    or helium) (see DICE Experiment referred in \cite{Grieder01}) :
     $\phi_{CR}(E_{CR} = 6\cdot 10^{15} eV)\simeq 9\cdot
     10^{-12}cm^{-2}s^{-1}$.
Within a Shower Cerenkov angle $\Delta\theta = 1^{o}$
     at a distance  $d =167 km \cdot \sqrt{\frac{h}{2.2 km}}$
     (zenith angle $\theta \simeq 87^{o}- 88^{o}$)
     giving  a  shower area $ [A = \pi \cdot(\Delta\theta \cdot d)^2\simeq 2.7 \cdot 10^{11} cm^2 (\frac{d}{167 km})^2 ]$,
the consequent event rate per night  for a Magic-like telescope
with a field of view of $[\Delta\Omega =(2^o \cdot 2^o)\pi \simeq
3.82 \cdot 10^{-3} sr.]$ is  \\
%%%% ($[\Delta(t)= 4.32 \cdot 10^4s]$):
\[N_{ev}=\phi_{CR}(E= 6\cdot 10^{15} eV)\cdot A \cdot \Delta \Omega
\cdot \Delta(t) \simeq 401/12 h \]
      Thus one may foresee that   nearly every
      two minutes  a horizontal hadronic  shower  may be observed by  Magic if it were pointed towards the horizon
      at zenith angle $87^o-88^o$.  Increasing the altitude $h$ of the observer, the
      horizon zenith angle grows: $\theta \simeq [90^o + 1.5^o
      \sqrt{\frac{h}{2.2km}}]$.
       In analogy at a more distant horizontal edges (standing at height $2.2
       km$ as for Magic, while observing at zenith angle $\theta \simeq 89^o- 91^o$
         still above the horizons) the observation range $d$ increases : $d= 167\sqrt{\frac{h}{2.2 km}} + 360 km = 527
       km$;  the consequent shower area widen by more than an order of
       magnitude (and more than  three order respect to vertical showers) and the foreseen event number,
       now for much harder CR at $E_{CR} \geq 3\cdot 10^{17} eV$,  becomes:
        $$N_{ev}=\phi_{CR}(E= 3\cdot 10^{17} eV)\cdot A \cdot \Delta \Omega \cdot
      \Delta(t) \simeq 1.6 /12 h$$
       Therefore at  $\theta \simeq 91.5^o$, once per night, a UHECR around EeV energies,
      may blaze  the Magic (or Hess,Veritas, telescopes).
        A long trail of secondary muons is associated to each of these far primary Cherenkov flash  in a very huge
        area.     These muons showering nearby the telescope, while they decay in flight
        into electrons, producing  tens-hundred GeVs  mini-gamma
        showers, is  also detectable at a rate discussed in the following section.

\section{ Muon's Arcs and Gamma  Flashes by $\mu^\pm \rightarrow e^\pm $}
    As already noted the  photons from a horizontal UHECR  may be also replenished  by  the secondary
    tens-hundred GeVs muons: they can  either  decay in flight as a gamma
    flashes,  or they may hit the telescope and their muon Cerenkov lights  "paint" arcs or rings  within the detector.
    Indeed these secondary very penetrating muon bundles
     may cover distances of hundreds km  ($\simeq 600 km \cdot\frac{E_{\mu}}{100\cdot GeV}$) away from the shower origin.
    To be more precise a part of the muon primary energy will be dissipated
     along their path in  air of $360$ km  (nearly a
      hundred GeV energy loss); thus a primary $130-150$ GeV muon is necessary to reach a final
       $30-50$ GeV energy at the minimal  Magic threshold value.
    Let us remind the characteristic multiplicity of secondary muons in a shower:
    $ N_\mu \simeq 3\cdot 10^5 \left( \frac{E_{CR}}{PeV}\right)^{0.85} $ \cite{Cronin2004} for GeV muons.  For the harder component (around 100
   GeV), the  muon number is  reduced almost inversely proportionally to energy
  $ N_\mu(10^2\cdot GeV) \simeq 1.3\cdot 10^4 \left( \frac{E_{CR}}{6 \cdot
  PeV}\right)^{0.85}$.
   These values must be compared to the larger peak multiplicity (but much lower energy) of
   electro-magnetic showers: $ N_{e^+ e^-} \simeq 2\cdot 10^7 \left(
   \frac{E_{CR}}{PeV}\right); N_{\gamma} \simeq  10^8 \left(\frac{E_{CR}}{PeV}\right) $.
    As mentioned before, most of the electromagnetic tail  is lost (exponentially) at
  horizons (for a slant depth larger than a few hundreds of
  $\frac{g}{cm^2}$), excluding re-born, upgoing $\tau$ air-showers
  \cite{Fargion2004},\cite{Fargion2004b} to be discussed later.
   Therefore   gamma-electron pairs are only partially  regenerated
    by the penetrating muon decay in flight, $\mu^\pm \rightarrow \gamma, e^\pm$
   as a parasite  electromagnetic showering \cite{Cillis2001}.
   Indeed $\mu^\pm $  may decay in flight (let say  at $100$ GeV energy,at $2-3\%$ level within a $12-18$ km distances)
    and they may inject more and more lights, to their primary (far born) shower beam.
    The ratio between $\gamma$, $\pm e$ over $\pm \mu$ offer a  clear hint of the
    Shower evolution \cite{Fargion2004b}.  These tens-hundred GeVs  horizontal muons and their associated mini-Cerenkov $\gamma$ showers are generated by:
    (1) either a single muon mostly produced at hundreds of kilometers  by a single  primary
    hadron  with an energy of  hundreds GeV-TeV; \\
        (2) rarer muons, part of a wider horizontal  bundle
    of large multiplicity born at TeVs-PeV  or higher energies, as secondary of an horizontal shower.
     Between the two cases there is a smooth link.
     A whole continuous spectrum  of multiplicity  begins from a unique muon up to a multi muon shower production.
     The  dominant noisy "single" muons at hundred-GeV energies
     will lose memory of their primary low energy and  hidden  mini-shower, (a hundreds GeV or TeVs hadrons );
      a single muon  will blaze just alone.
    The frequency of muon "single" rings or arcs  is larger than the muon bundles and it is based on solid observational data
    (\cite{Iori04} ; \cite{Grieder01}, as shown in Fig.2  see also the references on the  MUTRON experiment therein.
     The  event number due to the "noise" is:\\
     $$N_{ev- \mu} (\theta = 90^o)= \phi_{\mu}(E\simeq 10^{2} eV) \cdot A_{Magic} \cdot \Delta \Omega \cdot
      \Delta(t) \simeq 120 /12 h$$
      The additional gamma  mini-showers around the telescope due to a decay
         of those muons in flight (with a probability $p\simeq 0.02$),
         %recorded within a
         %larger collecting  Area $A_{\gamma} \geq 10^9 cm^2$
         is even a more frequent source of noise  (by a factor $\geq
         8$):\\
       $$N_{ev- \mu\rightarrow \gamma}\geq \phi_{\mu}(E\simeq 10^{2} eV)\cdot p \cdot A_{\gamma} \cdot \Delta \Omega \cdot
      \Delta(t) \simeq 960 /12 h$$
      These   single background gamma-showers must take place nearly once
       per minute (in a negligible hadronic background) and they represent a useful  tool
       to calibrate the possibility of detecting  Horizontal CR.

 \section{Showers  precursor and their late Muon tails}
    On the contrary PeVs (or higher energy) CR shower Cerenkov lights
     maybe  observed, more rarely, in coincidence  both by their primary
     and by their later secondary arc and gamma mini-shower.
   Their $30-100$ GeV  energetic muons are flying  nearly undeflected
  $\Delta \theta \leq 1.6^o \cdot \frac{100 \cdot GeV}{E_{\mu}}\frac{d}{300 km}$
  for a characteristic horizons distances d , partially bent by the geo-magnetic fields ($\sim$ $0.3$
  Gauss).  As mentioned, to flight   through the whole horizontal air column depth
  ($360$ km equivalent to $360$ water depth) the muon   lose nearly $100$ GeV; consequently the initial muon energy should be a little  above this threshold
   to be observed by Magic: (at least $ 130-150 $ GeV).   The deflection angle is  small:
    $\Delta \theta \leq 1^o \cdot \frac{150 \cdot GeV}{E_{\mu}}\frac{d}{300  km}$).
   Given the area of Magic ($A = 2.5 \cdot 10^6 cm^2$) we expect   roughly the
   following number of events due to  direct muons hitting the Telescope, flashing  as rings and arcs, each
   night:\\
  $$N_{ev}=\phi_{CR}(E= 6\cdot 10^{15} eV)\cdot N_\mu(10^2\cdot GeV) \cdot A_{Magic} \cdot \Delta \Omega \cdot
      \Delta(t) \simeq 45 /12 h$$ \\
to be correlated (at $11\%$ probability) with the above results of
$401$ primary Cerenkov flashes at the far distances.   Moreover,
the same muons are decaying in flight  at a minimal probability
$2\%$   leading to   mini-gamma-showers  in a  wider  area
($A_{\gamma}= 10^9 cm$). The related number of events we expect
   is:\\
   $$N_{ev}= \phi_{CR}(E= 6\cdot 10^{15} eV)\cdot N_\mu(10^2\cdot GeV) \cdot p \cdot A_{\gamma} \cdot \Delta \Omega \cdot
      \Delta(t) \simeq 360 /12 h$$\\      Therefore,  at $87^{\o}-88^{\o}$ zenith angle, there is a flow
      of a few dozens of    primary CR (at $E_{CR}\simeq 6\cdot 10^{16} eV$), whose earliest showers
         and consequent secondary muon-arcs as well as      nearby muon-electron mini-shower
         take place at comparable rate (one every  $120$ s).
         Therefore they may occur in time coincidence. Sometime
         also a bundle of arcs may better point to the C.R. Shower
         arrival. A well known  analogy occurs in meteor showers
         by their group perspective pointing at their origination.
       The time arrival of the prompt Shower respect the late muon one may be
          estimated by geometrical and relativistic arguments and
          may reach a detectable  delay value \cite{Cazon04}.
          These $certain$,  clustered, signals offer
          an unique tool for    calibrating Magic (as well as  Hess,Cangaroo,Veritas Cerenkov Telescope Arrays)
           for Horizontal High Energy Cosmic Ray Showers.      Some rarer events may contain at once both Rings,Arcs
            and tail    of gamma  shower as well as a Cerenkov  far primary shower.
                It is possible to estimate also the observable muon-electron-Cerenkov
                photons  from up-going albedo muons observed by the most
        recent ground experiments  \cite{NEVOD}  \cite{Decor}:
        their flux   is already suppressed at zenith angle $91^o$ by at least two orders of
         magnitude and by four orders for up-going zenith  angles $94^o$.
   Pairs or bundles are nevertheless rarer (up to $\phi_{\mu} \leq 3 \cdot 10^{-13} cm^{-2}s^{-1}sr^{-1}$
   \cite{NEVOD}  \cite{Decor}). They are never associated to up-going showers  (excluding the case of
   tau air-showers or  the Glashow $\bar{\nu_e}-e\rightarrow W^-$ and  the comparable $\chi^o + e\rightarrow
   \tilde{e}$).
%%%   detectable by stereoscopic Magic or Hess array telescopes, selecting and evaluating their column depth origination, just discussed below.
%---------------------------------------------
\section{Glashow $\bar{\nu_e} + e \rightarrow W^{-}$ Neutrino Astronomy }

  We note that UHE neutrinos may interact hitting nucleons and shower on air while crossing longest horizontal column
  depth ($\simeq 360 g cm^{-2}$) by charged and neutral currents;
   a defined competitive horizontal  events by UHE $6.3$ PeV, Glashow $\bar{\nu_e}-e\rightarrow W^- \rightarrow $
    (hadrons or electromagnetic showers)
   are also taking place;  one should remember that at Glashow's peak resonance
   the probability conversion is $\simeq 5 \cdot 10^{-3}$ at 360 km air distance and the consequent event number is:
   $$N_{ev}= \phi_{\bar{\nu_e}} (E= 6\cdot 10^{15} eV) \cdot A\cdot \Delta \Omega \cdot
      \Delta(t) \simeq  5.2 \cdot 10^{-4}/12 h$$
      assuming the minimal  GZK neutrino flux : $\phi_{\bar{\nu_e}} (E= 6\cdot 10^{15} eV)\simeq 5 \cdot
      10^{-15}$  $ cm^{-2} s^{-1}sr^{-1}$; the consequent GZK flat spectra
       energy fluency is assumed as : $\phi_{\bar{\nu_e}}\cdot E = 30 eV  cm^{-2} s^{-1}sr^{-1} $.
    A comparable number of events are also taking place smoothly at
    PeVs neutrino energies by nucleon charged current interactions (and marginally, by neutral
    current ones): ${\nu}_e + N \rightarrow e^- + X $,
${\nu}_{\mu} + N \rightarrow \mu + X $, ${\nu}_{\tau} + N
\rightarrow \tau + X $,
 and  ${\nu}_e + N \rightarrow {\nu}_e + X $, ${\nu}_{\mu} + N \rightarrow
{\nu}_{\mu} + X $, ${\nu}_{\tau} + N \rightarrow {\nu}_{\tau}+ X
$. Naturally the air column depth at $360$ km distance is too dim
to let the Cerenkov signal to survive the Magic threshold, but
more muons bundle and showering or additional stereoscopic array
(or wider telescope area as $10^3$ $m^2$ ones ) may still record
such a far shower. The full night year record of $60$ or more
Magic like telescope may reach a dozen of  events with a quite
 guaranteed $\bar{\nu_e}-e\rightarrow W^- \rightarrow $
    (hadrons or electromagnetic showers) channel.

\subsection{ SUSY $\chi^o + e\rightarrow \tilde{e} \rightarrow \chi^o +
e$ electromagnetic showers}

       Therefore   in a year of observations, provided that the data are taken at  night,
        assuming a minimal GZK flux,  a full crown array of
   a $90$ Magic-like telescopes set on a circle  ($2\cdot \pi = 360^o$) and facing the horizon,
   would give a number of events  comparable to a   $Km^3$  detector, (more than a dozen per year).
    Indeed  the present unique Magic Telescope
    pointing at the  horizon  offers a detection comparable to   the present
    AMANDA $\simeq  1\%  Km^3$ effective volume.
     It should be noted that up-going  $\bar{\nu_e}-e\rightarrow
     W^-$ and $\chi^o + e\rightarrow \tilde{e}$  below the
     horizons are possible within nearly $1^o$ because the Earth is not
     totally opaque at those small angles for these two tuned resonant
     interactions. The event number might be comparable with the
     $\bar{\nu_e}-e\rightarrow W^- \rightarrow $
    (hadrons or electromagnetic showers), provided that UHECR are
    originated in SUSY neutralino as UHE neutrinos in top-down
    models.

\section{Discovering UHE ${\nu_\tau}$ by Horizontal Tau Showers}

The appearance of horizontal UHE neutrino $\tau$ interaction in
matter   $\bar{\nu_\tau} +N $ ${\nu_\tau}+ N \rightarrow \tau^\pm$
 leading, in air, to air-showers (Hortaus or Earth-Skimming neutrinos)
    has been widely studied \cite{Fargion1999},\cite{Fargion
2002a},\cite{Bertou2002},\cite{Feng2002}; see also
\cite{Fargion03},\cite{Fargion2004},\cite{Jones04},
  \cite{Yoshida2004},\cite{Tseng03}and more recent \cite{Fargion2004b} ;
   Their rise from the Earth is source
   of rare clear signals for neutrino UHE astronomy .
     The higher the observer is located the wider is the view
     angle of the underneath Earth. The consequent up-going muon
     flux, by tau air-showers, are changing with quota \cite{Fargion2004b}.  The upward gamma and electron pair showering
may be also a very
   solid probe of the $\tau$ air-shower nearby showering
   \cite{Fargion2004b}; nevertheless the long lived muon bundle
   trace, correlated also with the Cerenkov flashes $below$ the horizons, may allow to
   confirm the UHE neutrino origination of the Horizontal blazing
   and its muon detection and arrival timing may better define its primary energy.

\subsection{Polarization Filters rejecting  reflected  Cerenkov Flash }
   One should  take into account the noise due to any inclined
   down-ward CR shower mirrored by the Sea leading to apparent
   up-going Cerenkov flashes; this mirror noise signal
   will be disentangled by the  co-presence of the
    twin up (above horizons primary Shower) and down image (reflected image)
     by the different muon versus Cerenkov
      arrival direction. An additional misleading signal may occur
   also by the muons bent by geomagnetic fields or by albedo scattering on the Earth.
   The very discriminant filter of
       fake mirrored Cerenkov showers is  offered by their extreme polarization
       imprinted   by the planar reflection (either sea or a ground).
       Therefore it must be considered a polarization filter of
       the arrival photons.

     \section{GRBs and UHECR correlated with UHE $\nu$ and Upward Shower }
   While the persistent  UHE gamma sources , as BL Lacs are a
   natural target for UHE neutrino Astronomy to follows, the rare GRBs are a more powerful, but
   rarer and unexpected  sources. However the fast follow up of GRB
   direction by Magic Telescope, may be fast enough to catch a PeV neutrino
   showering  at the horizons as well as an upward one from the Earth. Indeed the rare
    and sharp presence of a UHE neutrino scattering in air , by
    Glashow resonance at horizons, may be quite rare but above the
    noise at edge zenith $91.5^o-92.5^o$. The tau neutrinos may come at PeVs-EeVs energy from a much wider
    angle windows, and they are observable, in principle, from $91.5^o-120^o$ angle view.
     We may expect a few rare discover of GRBs or SGRs blazing from the dark horizons or
    from the Earth a year, if  a wide zenith angle windows will be realized.
      The same events might be searched in correlation with UHECR
      :therefore it might be interesting to search for UHECR event
      arrival directions that may contains UHE neutrinos in
      correlated time while arriving at Horizons or beneath the
      Earth. This might be possible because the fast tracking of
      Magic telescopes and the possibility to bend the mirror
      below to the Earth side. This may occur in correlated TeV
      activities of BL Lacs sources\cite{Fargion 2004d}.

      \section{ Conclusions}
      The Horizontal shower detection may be improved projecting a coexistence of a
      scintillator Crown Array (for muons and electromagnetic secondaries)
       aligned with Crown Magic Telescope Array; they will  able to verify the electromagnetic
      and muon shower nature in coincident arrival time with optical Cerenkov flashes.
        The Crown Array may encompass a large area
       as hundreds square meters array (see last figure) and
      its structure may be added around the Cerenkov Crown telescope
      array at the same mountain top. Its ability to discover muons
      (sometimes in correlation with the Cerenkov telescope)
      as well as the electromagnetic trace $\gamma, e^+,e^-$ nature and ratio, for a total area of
      63 $m^2$, is superior by a factor two respect a  (single ) Magic telescope, because it  enjoys a
      larger solid angle and a longer  recording time. Magic-Crown
      and Muon Crown array may reveal tens of thousands horizontal
      CR shower and a ten of up-going neutrino induced air-shower
      a year (for GZK minimal fluxes).  Finally the possible detection of up-going (a
      few-ten  GeV)  muons induced  neutrino hitting the Earth-Ground energetic
      Solar Flare at a rate $\phi_{\mu}\simeq 10^{-10} cm^{-2}s^{-1}$
      for a thousand of seconds, during the maximal solar flare activity,
    might also be a target of such future Muon-Cerenkov astronomy
    \cite{Solar} based on  largest telescopes looking on the nights down-ward the Earth toward the Sun.
     To conclude, while Magic looks
      upward to investigate the exciting  Low Gamma GeVs Astronomy,
    the same telescope looking at the horizon must discover higher
    (PeVs-EeVs) CR at high frequency, and rarely along the  edge,
       GZK $\bar{\nu_e}-e\rightarrow W^-$ air-showering neutrinos;
    some times looking downward also most power-full Solar flare, as the last 28th Oct. 2004 and  20th Jen. 2005 ones,
     may shine neutrinos whose up-going muons may be observable well below the horizons.

     In conclusions   new array of crown detectors might be better tuned for the horizons and  below the horizons
     seeking    $\nu_{\tau} \rightarrow \tau$,$\bar{\nu_e}-e\rightarrow W^-$ Earth-Skimmimng air-showers Astronomy
      and, surprisingly even  SUSY $\tilde{e}$ air-showering lights.

%%%%%%%%%%%%%%%%%%%%%%%%%%%%%%%%%%%%%%%%%%%%%%%%%%%%%%%%%%%%%%%%
\begin{figure}[h!]  %%% FIGURE 1 %%%
\epsfysize=5cm \hspace{4.0cm} \epsfbox{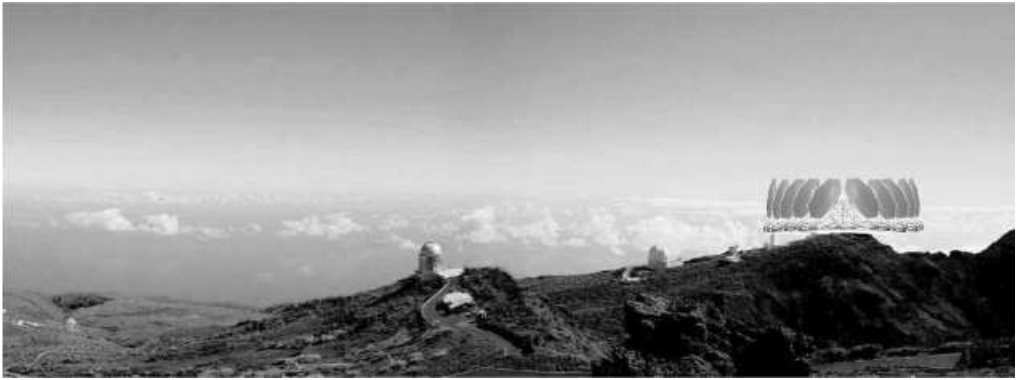}
\vspace{0.3cm} \caption[h]{The Earth view from Canarie sites for a
Magic-like Crown Telescope Array facing the Horizons, blazed, in
dark nights, by CR showers and rare up-going Tau Air-Shower
Cerenkov flashes; the  telescopes in circular array (of tens
telescopes at $360^o$ ) may test a wide area volume, nearly a
$km^3$ of air mass. The present (real) single Magic telescope is
located in the bottom left corner of the picture}
\end{figure}
\begin{figure}[h!]  %%% FIGURE 1 %%%
\epsfysize=7cm \hspace{4.0cm} \epsfbox{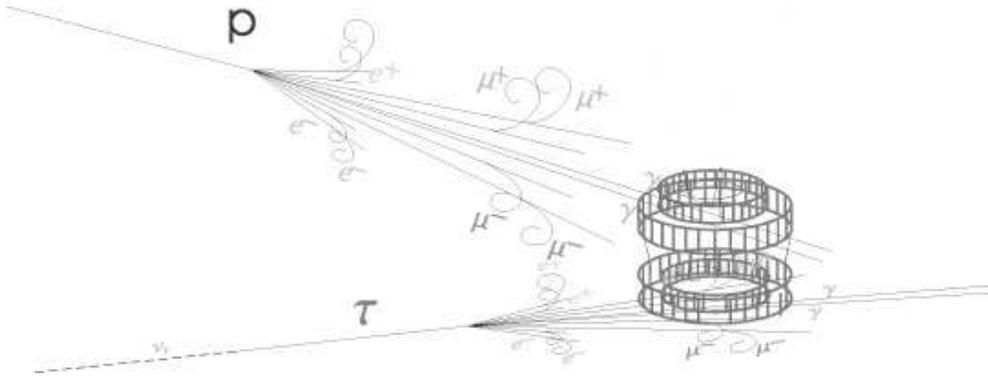}
\vspace{0.3cm} \caption[h]{A schematic figure of a Crown circular
detector array located in Balloon \cite{Fargion 2005} or on the
top of a mountain (possibly screened by a roof to avoid vertical
cosmic electro-magnetic C.R. noise) by a $\simeq 20-30$ meter
radius size, whose inner-outer aligned tiles are able to reveal,
by time of flight, the crossing and the azimuth directions of
horizontal muons (and electron pairs and gamma) bundles; in the
picture one observe a downward C.R. (a proton, label by p)and an
upward Horizontal Tau Air-Shower (label by $\tau$); the  $360^o$
tile disposal guarantees a wide azimuth solid angle view. The twin
(upper and lower) rings distance (height of tens or hundred meters
at the mountain top) is able to better disentangle the zenithal
arrival direction of the Horizontal Shower particles above or
below the Earth edges. A smaller twin scale device maybe dressed
around present AUGER detectors to improve their zenith angular
resolution to Horizons, in order to disentangle Tau  induced
showers from the Ande \cite{Fargion1999},\cite{Fargion
2002a}\cite{Bertou2002}.  }
\end{figure}
%%%%%%%%%%%%%%%%%%%%%%%%%%%%%%%%%%%%%%%%%%%%%%%%%%%%%%%%%%%%%%%%

\end{document}